# Supercontinuum generation in silicon Bragg grating waveguide

Neetesh Singh[1,2*], Manan Raval[1], Erich Ippen[1], Michael. R. Watts[1], and Franz X. Kärtner[1,2]
[1]Research Laboratory of Electronics, Massachusetts Institute of Technology, 77 Massachusetts Avenue, Cambridge, MA 02139, USA
[2]Centre for Free Electron Laser Science (CFEL)-DESY and University of Hamburg, Notkestrasse 85, 22607 Hamburg, Germany
*neeteshs@mit.edu

**Abstract:** Supercontinuum generation is an extensively studied and arguably the most important and all-encompassing nonlinear phenomenon. Yet, we do not have a good control over all the signals generated in this process. Usually a large part of an octave spanning spectrum has orders of magnitude too much weaker signal than the peak to be useful for any application. In this work we show strong signal generation within a supercontinuum using a silicon Bragg grating waveguide. We show up to 23 dB of signal enhancement over a 10 nm full-width-at-half-maximum bandwidth at the Bragg resonance in the telecom window. Since the grating is made by depositing charge carriers periodically, thus avoiding any dimensional change in the waveguide, it can allow other functionalities offered by the induced electric field, such as second harmonic generation. The ease of grating fabrication, whether with dimensional variation or doping, makes such a device useful for enhancing signal strength at any desired frequencies with high precision within a supercontinuum independent of material platform. We believe this work opens a new avenue for supercontinuum enhancement on demand in integrated photonics.

**Introduction:** A classic example of nonlinear photonics is the supercontinuum (SC) generation that has been well studied in waveguides and fibers for many years. By launching high power pulses into a nonlinear medium in the anomalous dispersion regime it is now routine to achieve octave spanning supercontinuum for a variety of applications. However, the supercontinuum is usually not flat, that is, many spectral windows within the SC, sometimes spanning 100s of nanometers, have very weak signal [1]. This is a problem for applications relying on the strength of the signals generated in a desired spectral window such as carrier envelope phase stabilization of a mode-locked laser, telecom/datacom and frequency synthesis [2-5]. Supercontinuum spectrum is usually not flat for a given optical mode because SC based on anomalous dispersion consists of dispersive waves and solitons that have fixed spectral locations determined by the dispersion profile of a waveguide defined by its cross-section. By varying the cross-section, thus the dispersion, along the length of a waveguide one can achieve much flatter SC by controlling the generation of the dispersive waves and the solitons accordingly along the length, thus providing some control over the wavelength of the nonlinearly generated signal [6]. However, the SC enhancement based on such non-uniform waveguides has weak spectral windows that can span over 10 nm, and the wavelengths of the dispersive waves are dependent on the pump power. This limitation especially becomes a significant problem for on-chip devices where the pump power tends to be limited.

To address the challenge of signal generation at desired spectral windows with sub-nm precision, an interesting approach is to modulate the group velocity dispersion and/or nonlinearity at the pump wavelength, for example, by periodically varying the waveguide width along the length thus giving rise to higher order dispersive waves [7-9]. However, this technique shows less flexibility in the spectral location and amplitude of the enhanced signal, as well as imposes fabrication challenges, for example, in a cladding modulated integrated waveguide [10].

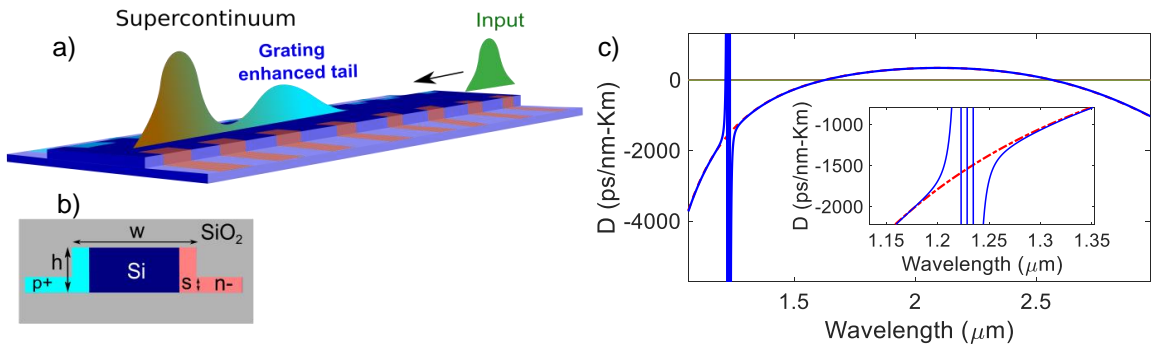

Fig.1 a) The schematic of the device with the input and the output pulse. b) The crossection of the waveguide and c) the dispersion of the waveguide with (solid blue) and without grating (red dash), zoomed in near the Bragg resonance (inset).

In this work, we demonstrate for the first time on-chip Bragg resonance based signal enhancement that can potentially be used for generating signal anywhere in a supercontinuum. The technique relies on the weak-grating-based strong modulation of waveguide dispersion near the Bragg resonance that can cover many nanometers for a grating having a bandwidth less than a 1nm [11]. Due to the precise control of the period of the grating, thus the Bragg resonance wavelength, this offers excellent control over the wavelength of the signal enhancement, independent of the pump power and hence, on-demand shaping of the spectrum anywhere in the supercontinuum. Here we show for the first time on a chip more than 20 dB enhancement over 10 nm of the bandwidth in the telecom window. Moreover, the grating is fabricated by depositing charge carriers periodically along the length of the waveguide, thus avoiding any dimensional change of the waveguide while also potentially offering other functionalities such as reducing the life time of the free carriers generated by the nonlinear absorption, and second harmonic generation through electric poling [12, 13]. Such a device will be useful for signal enhancement at any desired frequency with high precision within a supercontinuum independent of the material platform.

**Device design and results:** We designed a waveguide with width (w) of 800 nm, height (h) of 380 nm and slab thickness (s) of 65 nm as shown in Fig.1b. The grating is of higher order to serve the dual purpose of ease in fabrication and periodic poling of silicon for second harmonic generation [12, 13], which will be reported in the future work. The grating was formed by doping the waveguide with *n* and *p* type species overlapping the slab and partly the strip section of the waveguide by 30 nm, as shown in Fig. 1a and b. In a periodic structure like a Bragg grating the incident or forward propagating wave gets reflected by the Fresnel reflection at a wavelength dertermined by the periodicity of the index variation, thus giving rise to an overall reflected Bragg signal [14-16]. The strength of the coupling of the forward wave to the backward reflected wave is defined by the grating coupling constant, given as $k = \pi \delta n e / m \lambda_B$ where $\delta n$ is the modulation of the index of the grating, $e$ is the overlap of the mode of the incident wave with the grating, $m$ is the order (odd) of the grating and $\lambda_B$ is the Bragg wavelength which is equal to $2n_{eff}\Lambda/m$, where $n_{eff}$ is the average of the effective index of the mode at $\lambda_B$ with and without grating, and $\Lambda$ is the grating period which was designed to be 1650 nm. The $\delta n$ is ~2 to $4 \times 10^{-3}$, estimated using a mode solver and the emperical relation given in ref. [17]. Since we are using a weak grating ($\delta n \ll 0.1$), the grating modified total dispersion of the waveguide is calculated by using the propagation constant which is the sum of the propagation constant of the waveguide and the grating, as given in ref. [11,18]. The dispersion of the with- and without-grating waveguides for the transverse electric fundamental mode is shown in Fig. 1c. Here we see the presence of the grating causes sharp variation around the reflection wavelength, $\lambda_B$, due to the strong group delay dispersion introduced by the grating which is discussed later.

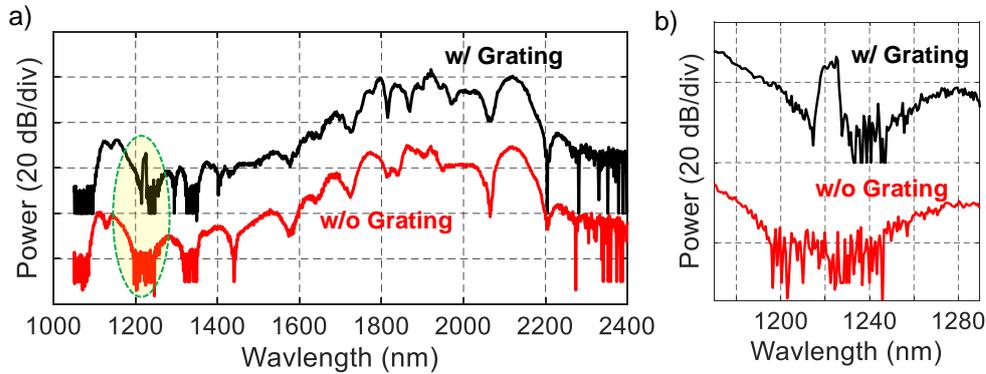

Fig.2 a) The measured supercontinuum spectra with (black) and without grating (red). b) Zoomed-in spectra near the grating response. The plots are shifted vertically for clarity.

The waveguide was fabricated in a standard 300-mm line CMOS foundry (CNSE SUNY polytechnic). To fabricate the grating, the dopants were ion implanted in the silicon layer to create *p*-type and *n*-type regions at a tilt angle of 20 degree to the normal of the wafer. In the experiment, the pump light at 1.95 μm was launched from an optical parametric oscillator having the pulse width of 250 fs at a repetition rate of 80 MHz, and was coupled into the waveguide with the peak power of 100-150 W using a free-space lens of numerical aperture ~ 0.6. Since the grating was based on charge carriers the propagation loss at the pump wavelength is expected to be 4-5 dB/cm [13]. The supercontinuum generation, based on the soliton fission process [19-24] is shown in Fig.2. Here we see the grating based signal enhancement of 23 dB with 10 nm of bandwidth at full width at half maximum (FWHM) around the Bragg wavelength of 1230 nm. There is a slight variation between the two SC spectra due to the coupled pump power variation and the propagation loss difference. The spectrum was taken with the resolution of 0.5nm. To demonstrate fine control on the grating-based signal enhancement, we fabricated gratings with the period varying in step size of 10 nm for which the measured supercontinuum is shown in Fig. 3. The enhanced signal is shifting 9 nm in wavelength for every 10 nm variation of the period of the higher order grating, thus illustrating a good control over the spectral location of the signal enhancement. The grating enhanced signal is broad and strong in the regions where the supercontinuum is weak and becomes narrower as the signal starts overlapping the stronger regions of supercontinuum (top plots of Fig.3). The signal strength increase varies 13-23 dB in the entire tested range with a respective bandwidth variation of 5–10 nm. The reduction in the strength of the grating enhanced signal in the upper plots is most likely caused by the weaker nonlinear interaction in the gratings owing to lower pump powers caused by loss in the longer waveguides used before the gratings due to fabrication constraints. The rest of the spectrum, above 1420 nm, is not shown as it varies negligibly from Fig. 2a. Also, we did not see any changes in the response of the enhanced signal with the application of up to 5 V reverse bias.

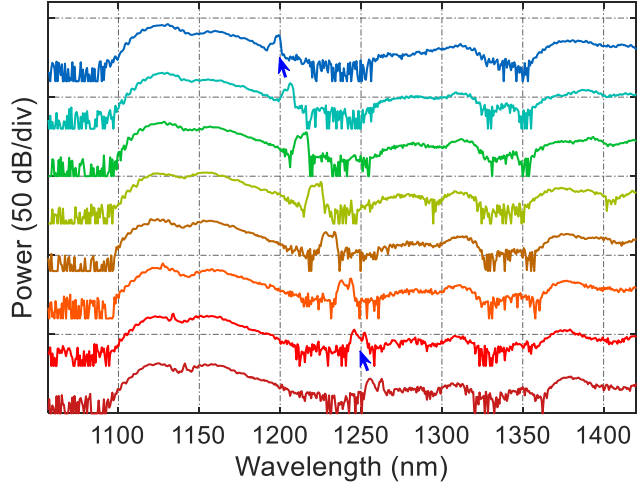

Fig.3. Measured supercontinuum spectra with grating period varying at 10 nm step size. The grating signals are indicated by the blue arrows shifting from 1260 nm up to 1200 nm. The plots are shifted vertically for clarity.

To understand the underlying mechanism, we solved the nonlinear Schrödinger equation (NLSE) using the split step Fourier method. To account for the fast variation afforded by the grating on the waveguide dispersion, we used the full dispersion of the waveguide, and spline interpolation to minimize the interpolation error, to extract the dispersion from the calculated effective index at discrete wavelengths. An NLSE similar to, $dE/dz = i(D_{w+g} + \gamma + i\alpha)E$, was solved to determine the signal enhancement. Here, $E$ is the pulse field envelop, $\gamma$ is the intensity dependent nonlinear term, $\alpha$ is the loss, and the dispersion term in frequency domain $D_{w+g} = \beta_{w+g}(\omega) - \beta_{w+g}(\omega_0) - (\omega-\omega_0) \beta_{1w+g}(\omega_0)$, where $\beta_{w+g}$ and $\beta_{1w+g}$ are the full dispersion and the first order derivative of the full dispersion of the grating waveguide (for detail see ref 11). In the simulation we used a 250 fs pulse, with the peak power of 200 W, the waveguide length was 5 mm, effective area of the pump mode was 0.3 µm$^2$, Kerr factor was $12\times10^{-18}$ m$^2$/W, and the propagation loss was 5 dB/cm. The simulated supercontinuum spectrum with- and without- grating is shown in Fig. 4.

In Fig. 4 we see the signal strength at the Bragg wavelength, $\lambda_B$ of 1230 nm, as in Fig. 2, is enhanced by 20 dB over that of the without-grating waveguide and the rest of the spectrum remains fairly similar. The group velocity dispersion without the grating is $1.24\times10^{-24}$ s$^2$/m near 1230 nm (normal dispersion), and with the grating the dispersion varies 4-5 orders of magnitude between normal and anomalous dispersion. Such a large grating-induced dispersion variation has been considered before for nonlinear pulse compression and dispersion compensation in optical fibers [25-27].

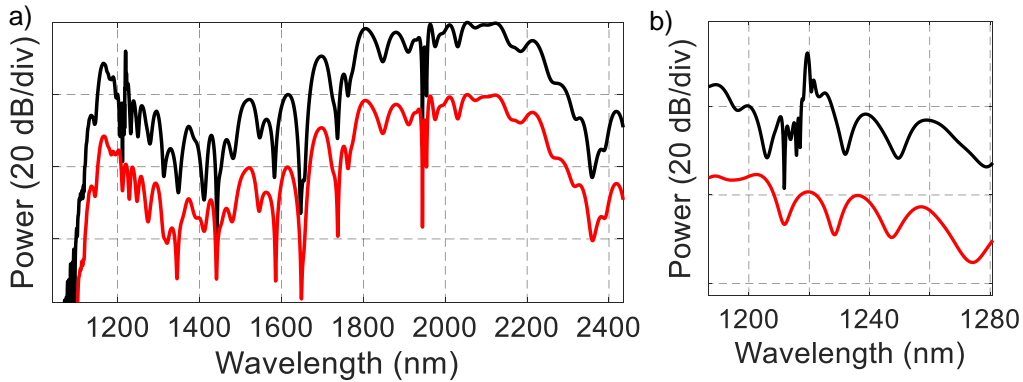

Fig.4 a). Simulated supercontinuum with (black) and without grating (red) with waveguide parameters similar to Fig.2. b) Zoomed-in spectra near the grating response. Plots are shifted vertically for clarity.

This extreme dispersion variation causes the signal around the Bragg resonance to be generated through Kerr nonlinearity [28-30]. In Fig. 5a,b we show how this grating-based dispersive wave is generated near the soliton fission point of roughly 2 mm. In the without-grating waveguide the original broadband dispersive wave around 1200 nm is generated. This gets strongly enhanced near the Bragg resonance in the grating waveguide and remains strong for the entire length of the waveguide. Although, we used a device of 5 mm length, a shorter device, slightly longer than 2 mm, can also generate efficient supercontinuum as the overall spectrum does not change beyond the soliton fission point as can be seen in Fig.5a,b.

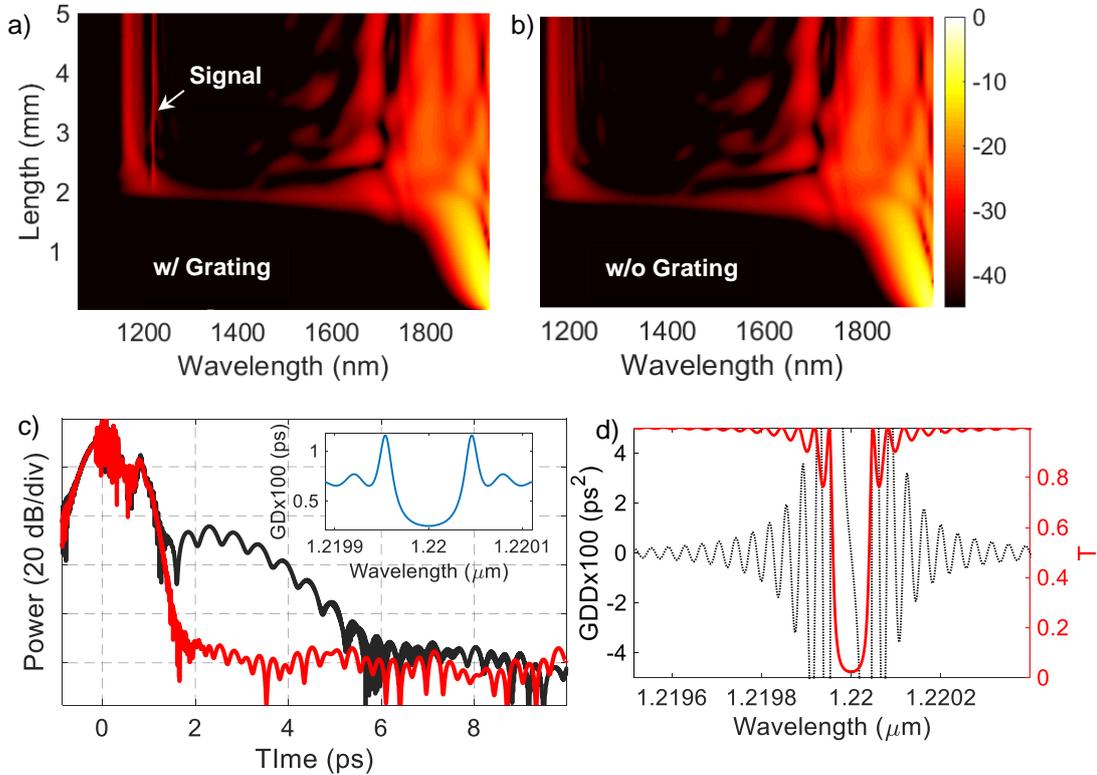

Fig. 5 a) and b) The spectrogram of the supercontinuum evolution with and without grating, respectively. c) The temporal profile of the pulse with (black) and without grating (red) at the output. A tail of the enhanced signal is seen spanning from 2 to 5 ps (black). The group delay (GD) for the grating is shown in the inset. d) The grating transmission (red) and the group delay dispersion (black dash).

In Fig. 5c we see the temporal evolution of the pulse at the output of the waveguide. The grating enhanced signal is lagging significantly behind the main pulse by several picoseconds because the grating adds extra group delay to these newly generated dispersive waves. The group delay introduced by the grating is shown in Fig. 5c (inset) where we see large spikes next to the reflection window which slowly oscillate down to the group delay introduced by the waveguide without grating which is 66 ps for a 5 mm long waveguide. These large oscillations in the group delay give rise to the rapidly varying normal and anomalous dispersion in the transmission region of the grating. The calculated transmission response of our grating and the group delay dispersion (GDD) is shown in Fig. 5d. Here the GDD oscillates between positive and negative values away from the resonance, which still imposes a strong variation of dispersion on the waveguide far away from the resonance thus giving rise to a broader grating-based signal enhancement outside the bandgap for a relatively narrowband grating. We must note here that a stronger grating can have high reflection which can cause adverse effect on the overall SC generation.

**Conclusion:** We have shown strong, grating-based signal enhancement of more than 20dB over a bandwidth of 10 nm (FWHM) in the supercontinuum generated in a silicon waveguide. Since the grating is based on charge-carrier induced index variation, the device can also be used for electrically poled second harmonic generation. Moreover, this technique can give precise control over the wavelength of the enhanced signal almost anywhere in the supercontinuum filling the spectral gaps present in supercontinuum with precision defined by the grating period which can be within nanometers. The enhanced signal is normally dispersed in time and can be compressed with an additional waveguide and, if required, be further enhanced by cross phase modulation [6].

**Acknowledgment:** This work was supported by Defense Advanced Research Projects Agency (DARPA) under the Direct on-chip digital optical synthesizer (DODOS) project—contract number HR0011-15-C-0056 and the DFG Priority Program SP2111 under contract PACE.

**Data availability:** The data that support the findings of this study are available from the corresponding author upon reasonable request.

**Author contribution:** N.S. conceived the project, conducted experiments and numerical analysis of the device. M.R. made the device lay-out for fabrication. E.I., M.R.W. and F.X.K. helped analyzing and supervising the experiment. All authors contributed in preparation of the manuscript.

**Competing interests:** The authors declare no competing interests.